\documentclass[11pt]{amsart}

\usepackage{fullpage}

\usepackage{amssymb, verbatim,url, algorithm, algorithmic}
\usepackage{appendix}
\usepackage{paralist}

\usepackage{amsmath}

\numberwithin{equation}{section}
\newtheorem{theorem}{Theorem}[section]

\newtheorem{definition}[theorem]{Definition}

\newcommand{\F}{\mathbb{F}}
\newcommand{\FB}{\mathcal{B}}

\newcommand{\G}{\mathbb{G}}

\newcommand{\bigO}{\mathcal{O}}
\newcommand{\n}{n^{\prime}}

\begin{document}

\title{Point Decomposition Problem in Binary Elliptic Curves}
\author{Koray Karabina}
\address{Dept. of Mathematical Sciences\\
        Florida Atlantic University\\
        Boca Raton, Florida, US}
\email{kkarabina@fau.edu}


\begin{abstract}
We analyze the point decomposition problem (PDP)
in binary elliptic curves. It is known that 
PDP in an elliptic curve group can be reduced to solving a particular system of multivariate non-linear system of equations
derived from the so called Semaev summation polynomials.
We modify the underlying system of equations by introducing some auxiliary variables.
We argue that the trade-off between lowering the degree of 
Semaev polynomials and increasing the number of variables is worth.
\end{abstract}




\maketitle



\section{Introduction}
\label{s: introduction}

\textit{Point decomposition problem} (PDP) in an additive abelian group
$\G$ with respect to a \textit{factor base} $\FB\subset \G$ is the following:
Given a point\footnote{We prefer to use \textit{point} rather than \textit{element} 
because elliptic curve group elements are commonly called points.} 
$R\in \G$, find $P_i\in \FB$ such that 
\[R=\sum_{i=1}^m{P_i}\]
for some positive integer $m$; or conclude that $R$ cannot be decomposed 
as a sum of points in $\FB$. 
\textit{Discrete logarithm problem} (DLP) in $\G$ with respect to a base $P\in \G$ 
is the following: Given $P$ and $Q=aP \in \G$ for some secret integer $a$, compute $a$. 
DLP can be solved using the \textit{index calculus algorithm} in two main steps. 
In the \textit{relation collection} step, fix a factor base $\FB$, and 
find a set of points $R_{i}=a_iP+b_iQ$ for some randomly chosen integers $a_i,b_i$, 
such that $R_i$ can be decomposed with respect to $\FB$, i.e., 
\[R_i=\sum_{j}{P_{ij}},\ P_{ij}\in \FB.\] 
Here, we may assume for convenience that  $P_{ij}$ are not necessarily distinct, and only finitely many of them are non-identity.
Note that each decomposition induces a modular linear dependence on the discrete logarithms
of $Q\in \G$ and $P_{ij}\in \FB$ with respect to the base $P$. 
After collecting sufficiently many relations\footnote{This is roughly when the number of relations exceeds $|\FB|$.},
\textit{linear algebra} step solves for the discrete logarithm of $Q\in \G$, as well as 
the discrete logarithms of the factor base elements.
Clearly, the success probability and the running time of the index calculus algorithm heavily depend on
the decomposition probability of a random element in $\G$, the cost of the decomposition step, and the size
of the factor base. In particular, the overall cost of the relation collection and the linear algebra steps
must be optimized with a non-trivial success probability.

In 2004, Semaev \cite{Semaev-04} showed that solving PDP in an elliptic curve group
is equivalent to solving a particular system of multivariate non-linear system of equations
derived from the so called \textit{Semaev summation polynomials}. Semaev's work triggered the possibility
of the existence of an index calculus type algorithm which is more efficient than the Pollard's rho algorithm
to solve the discrete logarithm problem in elliptic curves defined over $\F_{q^n}$,
which we denote ECDLP($q,n$). 
Note that Pollard's rho algorithm is a general purpose algorithm
that solves DLP in a group $\G$, and runs in time $\bigO(\sqrt{|\G|})$. 
Gaudry \cite{Gaudry-09} showed that Semaev summation polynomials can be effectively used
to solve ECDLP$(q,n)$ in heuristic time $\bigO(q^{2-\frac{2}{n}})$,
where the constant in $\bigO(\cdot)$ is exponential in $n$. For example,
Gaudry's algorithm and Pollard's rho algorithm solve ECDLP($q,3$)
in time $\bigO(q^{1.33})$ and $\bigO(q^{1.5})$, respectively. 
Due to the exponential in $n$ constant in the running time
of Gaudry's algorithm, his attack is expected to be more effective than
Pollard's rho algorithm if $n\ge 3$ is relatively small and $q$ is large.
Diem \cite{Diem-13} rigorously showed that ECDLP($q,n$) can be solved in an expected subexponential time
when $a(\log q)^{\alpha} \le n \le b(\log q)^{\beta}$ for some $a,b, \alpha, \beta >0$.
On the other hand, Diem's method has expected exponential running time $\bigO(e^{n(\log n)^{1/2}})$
for solving ECDLP($2,n$). As a result, index calculus type algorithms presented 
in \cite{Gaudry-09,Diem-13} do not yield ECDLP solvers which
are more effective than Pollard's rho method when $q=2$ and $n$ is prime.
The ideas for choosing an appropriate factor base in \cite{Diem-13}
have been adapted in \cite{FPPR-12,PQ-12}, 
and the complexity of the relation collection step
have been analyzed. 
In both papers \cite{FPPR-12} and \cite{PQ-12},
a positive integer $m$, which we call the \textit{decomposition constant}, is fixed to represent
the number of points in the decomposition of a random point in the relation collection step.
The factor base consists of elliptic curve points
whose $x$-coordinates belong to an $\n$-dimensional subspace $V\subset \F_{2^n}$ over $\F_2$, 
where $\n$ is chosen such that $m \n \approx n$.
We refer to PDP in this setting by PDP($n,m,\n$) throughout the rest of this paper.

Faug\`{e}re~et al.\ \cite{FPPR-12} 
showed, under a certain assumption, that
ECDLP($2,n$) can be solved in 
time $\bigO(2^{wn/2})$, where $2.376 \le w \le 3$ is the linear algebra constant.
The running time analysis in \cite{FPPR-12} considers the linearization technique
to solve the multivariate nonlinear system of equations which are derived
from the ($m+1$)'st Semaev polynomial $S_{m+1}$ during the relation collection step to solve PDP($n,m,\n$).
Faug\`{e}re~et al.\ further argue that, Groebner basis techniques may improve 
the running time by a factor $m$ in the exponent, where $m$ is the decomposition constant.
This last claim has been confirmed in the experiments in \cite{FPPR-12} for elliptic curves
defined over $\F_{2^n}$ with $n\in \{41,67,97,131\}$ and $m=2$.
Petit and Quisquater's heuristic analysis in \cite{PQ-12} claims that 
ECDLP($2,n$) can asymptotically be solved in time $\bigO(2^{cn^{2/3}\log n})$ for some constant $0<c<2$.
The subexponential running time in \cite{PQ-12} is based on a rather strong assumption 
on the behavior of the systems of equations that arise from
Semaev polynomials. In particular, it is assumed in \cite{PQ-12} that 
the degree of regularity $D_{\mathsf{reg}}$ and the first fall degree $D_{\mathsf{FirstFall}}$ of the underlying polynomial systems to solve PDP($n,m,\n$) are approximately equal.
The analysis in \cite{PQ-12} also assumes that $\n = n^\alpha$ and $m = n^{1-\alpha}$ for some positive constant $\alpha$.
Experiments with a very limited set of parameters $(n,m,\n)$, $n\in\{11,17\}$, $m\in\{2,3\}$, $\n=\left\lceil  n/m \right\rceil$
were conducted in \cite{PQ-12} in the favor of their assumption.

A recent paper by Shantz and Teske \cite{Teske-13}
presented some extended experimental results on solving PDP($n,m,\n$) 
for the same setting as in the Petit and Quisquater's paper \cite{PQ-12}. 
In particular, \cite{Teske-13} validates the degree of regularity assumption
in \cite{PQ-12} for the set of parameters 
$(n,m,\n)$ such that $n \in \{11,13,15,17,19,23,29\}$, $m=2$, $\n = \left\lceil n/m \right\rceil$; 
and for $(n,m,\n)$ such that $n \in \{11,13,15,17,19,21\}$, $m=3$, $\n = \left\lceil n/m \right\rceil$.
Shantz and Teske \cite{Teske-13} were able to extend their experimental data
for the parameters $(n, m, \n, \Delta)$, $n\le 48$, $m=2$, and where $\Delta=n-m\n$ is chosen
appropriately to possibly improve the running time of ECDLP($2,n$).  
In another recent paper \cite{Petit-13}, Yun-Ju~et al.\ exploit the symmetry in Semaev polynomials, 
and improve on the running time and memory requirements of the PDP($n,m,\n$) solver in \cite{FPPR-12}.
The efficiency of the method in \cite{Petit-13} is tested for parameters
$(n,m,\n)$ such that $n \le 53$, $m=3$, $\n = 3,4,5,6$.

Petit and Quisquater's heuristic analysis \cite{PQ-12} claims that index calculus methods 
for solving ECDLP($2,n$) is more effective than the Pollard's rho method
for $n>2000$, $m\ge 4$ and $m\n \approx n$.
However, all the experiments reported so far on solving PDP($n,m,\n$) for the set of parameters
$(n,m,\n, \Delta)$ with $\Delta = n-m\n \le 1$ and $m=3$  
are limited to $n\le 19$; see \cite{Teske-13, Petit-13}. 
Similarly, all the experiments for the set of parameters
$(n,m,\n, \Delta)$ with $m=3$ are limited to $\n\le 6$, 
which forces $\Delta \ge 2$ for $n\ge 20$. 
In general, it is desired to have $\n$ increasing as a 
function of $n$, rather than having some upper bound on $\n$, so that $n\approx m\n$
as assumed in the running time analysis of ECDLP($2,n$) solvers in \cite{FPPR-12,PQ-12}.
Therefore, it remains as a challenge to run experiments
on an extensive set of parameters $(n,m,\n)$ with larger prime $n$ values, $m\ge 4$, and $m\n \approx n$.
For example, it is stated in \cite[Section 4.1]{Petit-13} that
\begin{quote}
On the other hand, the method appears unpractical for $m=4$ even for
very small values of $n$ because of the exponential increase with $m$ of the
degrees in Semaev's polynomials. 
\end{quote}

In a more recent paper \cite{Galbraith-14}, Galbraith and Gebregiyorgis
introduce a new choice of variables and a new choice of factor base,
and they are able to solve PDP with various $n\ge 17$, $m=4$, $\n=3,4$
using Groebner basis algorithms;
and also with various $n\ge 17$, $m=4$, $\n\le 7$ using SAT solvers.

In this paper, we modify the system of equations, that are derived
from Semaev polynomials, by introducing some auxiliary variables.
We show that PDP($n,m,\n$) can be solved
by finding a solution to a system of equations derived from several 
third Semaev polynomials $S_3$ each of which has at most three variables. 
For a comparison, 
PDP($n,m,\n$) in $E(\F_{2^n})$ with decomposition constant $m=5$
would be traditionally attacked via considering
the Semaev polynomial $S_6$ with $5$ variables,
which is likely to have a root in $V^5$, 
where $V\subset \F_{2^n}$ is a subspace of dimension $ \n = \left\lfloor n/5 \right\rfloor$.
On the other hand, when $m=5$, our polynomial system consists of third Semaev polynomials $S_{3,i}$ ($i=1,2,3,4$),
and a total of $8$ variables which is likely to have a root in $V^5\times \F_{2^n}^3$,
where $V\subset \F_{2^n}$ is a subspace of dimension $\left\lfloor n/5 \right\rfloor$.
As a result, our technique overcomes the difficulty of dealing with the ($m+1$)'st Semaev polynomial $S_{m+1}$
when solving PDP($n,m,\n$) with $m\ge 4$. We should emphasize that choosing $m\ge 4$ is desirable
for an index calculus based ECDLP($2,n$) solver to be more effective than a generic DLP solver such as Pollard's rho algorithm. 
Our method introduces an overhead of introducing some auxiliary variables.
However, we argue that the trade-off between lowering the degree of 
Semaev polynomials and increasing the number of variables is worth.
In particular, we present some experimental results
on solving PDP($n,m,\n$) for the following parameters:
\begin{itemize}[]
\item [--] $n\le 19$, $m=4,5$, and $\n = \left\lfloor n/m \right\rfloor$. We are not aware of any previous experimental data for $n>15$ and $m=5$.
\item [--] $n\le 26$, $m=3$, $\n = \left\lfloor n/m \right\rfloor$. We are not aware of any previous experimental data for $n > 21$, $m=3$, and $\Delta = n -m\n \le 2$.
\end{itemize}
We observe in our experiments that regularity degrees
of the underlying systems are relatively low.
We also observe that running time and memory requirement of algorithms can be improved significantly if the
the Groebner basis computations are first performed on a subset of polynomials
and if the $\mathsf{ReductionHeuristic}$ parameter in Magma is set to be a small number;
see Section~\ref{s: Extend}. 
We are able to solve PDP($15,5,3$) instances in about 7 seconds (with 256 MB memory).
Note that, PDP($15,5,3$) is solved in about 175 seconds (with 2635 MB memory) in \cite{Semaev-15}. 
Our experimental findings with $m=3,4,5$ extend and improve on recently reported results
in \cite{Teske-13,Petit-13,Semaev-15}.

The rest of this paper is organized as follows. 
In Section~\ref{s: Semaev}, we recall Semaev polynomials
and their application to ECDLP($2,n$). In Section~\ref{s: Method},
we describe and analyze a new method to solve PDP($n,m,\n$) in $E(\F_{2^n})$.
In Section~\ref{s: Experiments}, we present our experimental
results. In Section~\ref{s: Extend}, we extend our results from Section~\ref{s: Method}.


\paragraph{\bf Acknowledgment}
The author of this paper would like to acknowledge two recent papers \cite{Semaev-15, Kosters-15}. 
Semaev \cite{Semaev-15} claims a new complexity bound $2^{c(\sqrt{n\ln n}})$
for solving ECDLP($2,n$) under the assumption that the degree of regularity in Groebner computations
of particular polynomial systems is $D_{\mathsf{reg}}\le 4$. Semaev also shows that
ECDLP($2,n$) can be solved in time 
$2^{o(\sqrt{n\ln n})}$ under a weaker assumption that $D_{\mathsf{reg}} = o(\sqrt{n/\ln n})$
The techniques used in \cite{Semaev-15} and in this paper are similar.
In \cite{Kosters-15}, Kosters and Yeo provide experimental evidence that
the degree of regularity of the underlying polynomial systems is likely to increase
as a function of $n$, whence the conjecture $D_{\mathsf{reg}} \approx D_{\mathsf{FirstFall}}$
may be false.

\section{Semaev Polynomials and ECDLP}
\label{s: Semaev}

Let $\F_{2^n} = \F_{2}[\sigma]/\langle f(\sigma)\rangle$ be a finite field with $2^n$ elements, 
where $f(\sigma)$ is a monic irreducible polynomial of degree-$n$ over the field $\F_2=\{0,1\}$.
let $E$ be a non-singular elliptic curve defined by the short Weierstrass equation
\[E/\F_{2^n}:\ y^2 + xy = x^3 + ax^2 + b,\ a,b\in \F_{2^n}.\]
We denote the identity element of $E$ by $\infty$.
The $i$'th Semaev polynomial associated with $E$ is defined as follows:
\begin{align}
S_i(x_1,x_2,\ldots,x_i) = \begin{cases} (x_1x_2+x_1x_3+x_2x_3)^2+x_1x_2x_3 + b & \mbox{if } i=3 \\
\mbox{Res}_X(S_{i-j}(x_1,\ldots, x_{i-j-1}, X), S_{j+2}(x_{i-j},\ldots, x_i,X)) & \mbox{if }i\ge 4, \end{cases}
\end{align}
where $1\le j\le i-3$.

Let 
\[V = \{a_0+a_1\sigma + \cdots + a_{n^\prime-1}\sigma^{n^\prime-1}:\ a_i\in \F_2,\  n^\prime\le n\}\subset \F_{2^n}\]
and define the factor base
\[\FB = \{P=(x,y)\in E:\ x\in V\}.\]
Recall that in PDP($n,m,\n$), we are looking for $P_i=(x_i,y_i)\in \FB$ such that
\begin{align}
\label{s: Decomposition}
P_1+\cdots P_m = R,
\end{align}
for some given point $R=(x_R,y_R)\in E$.
We refer to (\ref{s: Decomposition}) as an
$m$-decomposition of $R$ in $\FB$. We expect that, on average,
a random point $R\in E$ has an $m$-decomposition in $\FB$
with probability $2^{m\n}/2^n m!$ simply because
$| \FB | \approx 2^{\n}$ and permuting $P_i$ does not change the sum
$\sum P_i$ (see \cite{Gaudry-09}). As described in Section~\ref{s: introduction}, 
DLP in $E$ can be solved via an index-calculus based approach by computing
about $|\FB|$ explicit $m$-decompositions and solving a sparse linear system of about $|\FB|$ equations.
Therefore, the cost of solving ECDLP($2,n$) may be estimated as
\begin{align}
\label{s: cost ECDLP}
2^{\n}\frac{2^n m!}{2^{m \n}}C_{n,m,\n} + 2^{w^\prime \n},
\end{align}
where $C_{n,m,\n}$ is the cost of solving PDP$(n,m,\n)$, and
$w^\prime=2$ is the sparse linear algebra constant.
Semaev \cite{Semaev-04} showed that a decomposition of the form
(\ref{s: Decomposition}) exists if and only if the $x$-coordinates of $P_i$ and $R$ are 
zeros of the $(m+1)$'st Semaev polynomial, that is, $S_{m+1}(x_1,\ldots,x_m,x_R)=0$.
In the rest of this paper, we focus on solving  PDP$(n,m,\n)$ (and estimating $C_{n,m,\n}$)
via modifying the equation reduced by $S_{m+1}$.

\section{A new approach to solve point decomposition problem}
\label{s: Method}
Let $E/\F_{2^n}$, $V$, and $\FB$ be as defined in Section~\ref{s: Semaev}. Recall that
an $m$-decomposition of a point 
\[R = P_1+\cdots P_m,\] 
where $R=(x_R,y_R)\in E$, $P_i=(x_i,y_i)\in \FB$, 
can be computed (if exists) by identifying a tuple $(x_1,\ldots,x_m)\in V^{m}$ that satisfies
\begin{align}
\label{s: Semaev-m}
S_{m+1}(x_1,\ldots,x_m,x_R)=0
\end{align}
Note that $x_i$ belong to an $\n$-dimensional subspace of $\F_{2^n}$.
Therefore, (\ref{s: Semaev-m}) defines a system $\mathsf{Sys}_1$ 
of a single equation
over $\F_{2^n}$ in $m$ variables. 
In \cite{FPPR-12,PQ-12}, the Weil descent technique is applied, and a second system
$\mathsf{Sys}_2$ of $n$ equations over $\F_2$ in $m \n$ boolean variables is derived 
from $\mathsf{Sys}_1$. The cost $C_{n,m,\n}$ of solving PDP$(n,m,\n)$ in \cite{FPPR-12,PQ-12} is estimated
through the analysis of solving $\mathsf{Sys}_2$ using linearization and Groebner basis techniques.
Next, we describe a new approach to derive another system $\mathsf{Sys}_3$ of 
boolean equations such that a solution of $\mathsf{Sys}_3$ yields an $m$-decomposition
of a point $R$.

\paragraph{\bf{Notation}} Throughout the rest of this paper, we distinguish 
between two classes Semaev polynomials. The first class of Semaev polynomials
is denoted by $S_{m,1}(x_1,\ldots,x_m)$, which represents the $m$'th Semaev
polynomial with $m$ variables. The second class of Semaev polynomials
is denoted by $S_{m,2}(x_1,\ldots,x_{m-1}, x_R)$, which represents the $m$'th Semaev
polynomial with $m-1$ variables (i.e., the last variable $x_m$ is evaluated at a number $x_R$).

\subsection{The case: $\boldsymbol{m=3}$}
\label{s: m3}

Let $R=(x_R,y_R)\in E$. Notice that there exist $P_i\in \FB$ such that
\[P_1+P_2+P_3 - R = \infty\]
if and only if there exist $P_i\in \FB$ and $P_{12}\in E$ such that
\begin{align}
\begin{cases}
P_1+P_2-P_{12} = \infty\\
P_3+P_{12}-R = \infty
\end{cases}
\end{align}
Therefore, a $3$-decomposition of $R=P_1+P_2+P_3$ may be found as follows:
\begin{enumerate}
\item Define the following system of equations derived from Semaev polynomials
\begin{align}
\label{s: System-3}
\begin{cases}
S_{3,1}(x_1,x_2,x_{12}) = 0\\
S_{3,2}(x_3,x_{12},x_R)= 0.
\end{cases}
\end{align}
Note that this system is defined over $\F_{2^n}$ and has $4$ variables $x_1,x_2,x_3,x_{12}$.
\item Introduce boolean variables $x_{i,j}$ such that 
\[x_i = \sum_{j=0}^{\n -1}x_{i,j}\sigma^j,\] 
for $i=1,2,3$, and
\[x_{12} = \sum_{j=0}^{n}x_{12,j}\sigma^j.\] 
Apply the Weil descent technique to (\ref{s: System-3}) and define an equivalent system of $2n$ 
equations over $\F_2$ with $3\n + n$ boolean variables 
\[\{x_{i,j}:\ i=1,2,3,\ j=0,\ldots\n -1\}\cup\{x_{12,j}:\ j=0,\ldots n-1\}.\]
Solve this new system of boolean equations and recover $x_1,x_2,x_3 \in \F_{2^n}$ from $x_{i,j}\in \F_2$.
\end{enumerate}

Note that the proposed method solves a system of $2n$ 
equations over $\F_2$ with $3\n + n$ boolean variables rather
than solving a system of $n$ 
equations over $\F_2$ with $3\n$ boolean variables.

\subsection{The case: $\boldsymbol{m=4}$}
\label{s: m4}

Let $R=(x_R,y_R)\in E$. Notice that there exist $P_i\in \FB$ such that
\[P_1+P_2+P_3+P_4 - R = \infty\]
if and only if there exist $P_i\in \FB$ and $P_{12}\in E$ such that
\begin{align}
\begin{cases}
P_1+P_2-P_{12} = \infty\\
P_3+P_4+P_{12}-R = \infty
\end{cases}
\end{align}
Therefore, a $4$-decomposition of $R=P_1+P_2+P_3+P_4$ may be found as follows:
\begin{enumerate}
\item Define the following system of equations derived from Semaev polynomials
\begin{align}
\label{s: System-4}
\begin{cases}
S_{3,1}(x_1, x_2, x_{12}) = 0\\
S_{4,2}(x_3, x_4, x_{12}, x_R)= 0
\end{cases}
\end{align}
Note that this system is defined over $\F_{2^n}$ and has $5$ variables $x_1,x_2,x_3,x_4,x_{12}$.
\item Introduce boolean variables $x_{i,j}$ such that 
\[x_i = \sum_{j=0}^{\n -1}x_{i,j}\sigma^j,\] 
for $i=1,2,3,4$, and
\[x_{12} = \sum_{j=0}^{n}x_{i,j}\sigma^j.\] 
Apply the Weil descent technique to (\ref{s: System-4}) and define an equivalent system of $2n$ 
equations over $\F_2$ with $4\n + n$ boolean variables 
\[\{x_{i,j}:\ i=1,2,3,4\ j=0,\ldots\n -1\}\cup\{x_{12,j}:\  j=0,\ldots n-1\}.\]
Solve this new system of boolean equations and recover $x_1,x_2,x_3,x_4 \in \F_{2^n}$ from $x_{i,j}\in \F_2$.
\end{enumerate}

Note that the proposed method solves a system of $2n$ 
equations over $\F_2$ with $4\n + n$ boolean variables rather
than solving a system of $n$ 
equations over $\F_2$ with $4\n$ boolean variables.

\subsection{The case: $\boldsymbol{m=5}$}
\label{s: m5}

Let $R=(x_R,y_R)\in E$. Notice that there exist $P_i\in \FB$ such that
\[P_1+P_2+P_3+P_4+P_5 - R = \infty\]
if and only if there exist $P_i\in \FB$ and $P_{123}\in E$ such that
\begin{align}
\begin{cases}
P_1+P_2+P_3-P_{123} = \infty\\
P_4+P_5+P_{123}-R = \infty
\end{cases}
\end{align}
Therefore, a $5$-decomposition of $R=P_1+P_2+P_3+P_4+P_5$ may be found as follows:
\begin{enumerate}
\item Define the following system of equations derived from Semaev polynomials
\begin{align}
\label{s: System-5}
\begin{cases}
S_{4,1}(x_1,x_2,x_3,x_{123}) = 0\\
S_{4,2}(x_4,x_5,x_{123},x_{R})= 0
\end{cases}
\end{align}
Note that this system is defined over $\F_{2^n}$ and has $6$ variables $x_1,x_2,x_3,x_4,x_5,x_{123}$.
\item Introduce boolean variables $x_{i,j}$ such that 
\[x_i = \sum_{j=0}^{\n -1}x_{i,j}\sigma^j,\] 
for $i=1,2,3,4,5$, and
\[x_{123} = \sum_{j=0}^{n}x_{123,j}\sigma^j.\]
Apply the Weil descent technique to (\ref{s: System-5}) and define an equivalent system of $2n$ 
equations over $\F_2$ with $5\n + n$ boolean variables 
\[\{x_{i,j}:\ i=1,2,3,4,5\ j=0,\ldots\n -1\}\cup\{x_{123,j}:\  j=0,\ldots n-1\}.\]
Solve this new system of boolean equations and recover $x_1,x_2,x_3,x_4,x_5 \in \F_{2^n}$ from $x_{i,j}\in \F_2$.
\end{enumerate}

Note that the proposed method solves a system of $2n$ 
equations over $\F_2$ with $5\n + n$ boolean variables rather
than solving a system of $n$ 
equations over $\F_2$ with $5\n$ boolean variables.

\subsection{Analysis of new polynomial systems}
\label{s: Analysis}
One of the methods to solve a multivariate non-linear
system of equations is to compute the Groebner basis of the
underlying ideal. Groebner basis computations can be performed
using Faug\`{e}re's algorithms \cite{F4,F5}, which reduce the problem
to Gaussian elimination of Macaulay-type matrices $M_d$
of degree $d$. The Macaulay matrix $M_d$ encodes degree (at most) $d$
polynomials, that are generated during Groebner basis computation.
Therefore, the cost of solving a system of equations is determined by
the maximal degree $D$ (also known as the degree of regularity of the system)
reached during the computation. If $N$ is the number of variables in the system, 
then the cost is typically estimated as $O\left({N+D-1 \choose D}^w\right)$,
where ${N+D-1 \choose D}$ is the maximum number of columns in $M_D$
and $w$ is the linear algebra constant. In general, it is hard to estimate 
$D$. In the recent paper \cite{PQ-12}, it is conjectured that
the degree of regularity $D_{\mathsf{reg}}$ of systems arising from PDP($n,m,\n$) satisfies
$D_{\mathsf{reg}} = D_{\mathsf{FirstFall}} + o(1)$,
where $D_{\mathsf{FirstFall}}$ is the first fall degree of the system and defined as 
follows.
\begin{definition}\cite{PQ-12}
Let $R$ be a polynomial ring over a field $K$. Let $F:=\{f_1,\ldots,f_\ell\}\subset R$
be a set of polynomials of degrees at most $D_{\mathsf{FirstFall}}$. The first
fall degree of $F$ is the smallest degree $D_{\mathsf{FirstFall}}$ such that
there exist polynomials $g_i\in R$ with 
$\max_{i}{\deg(f_i)+\deg(g_i)} = D_{\mathsf{FirstFall}}$, satisfying
$\deg(\sum_{i=1}^{\ell}{g_if_i})<D_{\mathsf{FirstFall}}$ but 
$\sum_{i=1}^{\ell}{g_if_i}\ne 0$.
\end{definition} 
Experimental studies in recent papers \cite{PQ-12,Teske-13}
give supporting evidence that $D_{\mathsf{reg}} \approx D_{\mathsf{FirstFall}}$.
However, experimental data is yet very limited (see Section~\ref{s: introduction}) 
to verify this conjecture. In this section, we compute the first fall degree of the systems
proposed in Section~\ref{s: m3}, Section~\ref{s: m4}, and Section~\ref{s: m5}. 
Our experimental results in Section~\ref{s: Experiments} indicate that
$D_{\mathsf{reg}}\approx  D_{\mathsf{FirstFall}}$.

\paragraph{\bf $\boldsymbol {D_{\mathsf{FirstFall}}}$ of the system when $\boldsymbol {m=3}$} 
In this case, one needs to solve the system of $2n$ 
equations over $\F_2$ with $3\n + n$ boolean variables.
The system of equations is derived by applying Weil descent
to (\ref{s: System-3}) that consists of two Semaev polynomials $S_{3,1}$ and $S_{3,2}$. 
The monomial set of $S_{3,1}(x_1,x_2,x_{12})$ is
\[\{1, x_1^2x_2^2, x_1^2x_{12}^2, x_2^2x_{12}^2, x_1x_2x_{12}\}.\]
Therefore, the Weil descent of  $S_{3,1}(x_1,x_2,x_{12})$ yields a 
$2\n+n$ variable polynomial set $\{f_i\}$ over $\F_2$
such that $\max_i(\deg(f_{i})) = 3$. On the other hand,
the monomial set of $x_1\cdot S_{3,1}(x_1,x_2,x_{12})$ is
\[\{x_1, x_1^3x_2^2, x_1^3x_{12}^2, x_2^2x_{12}^2, x_1^2x_2x_{12}\}.\]
Therefore, the Weil descent of  $x_1\cdot S_{3,1}(x_1,x_2,x_{12})$ yields a 
polynomial set $\{F_i\}$ over $\F_2$
such that $\max_i(\deg(F_{i})) = 3$. It follows from the definition
that $D_{\mathsf{FirstFall}}(S_{3,1})\le 4$ because the maximum
degree of polynomials obtained from the Weil descent of $x_1$ is $1$. 
Similarly, the monomial set of $S_{3,2}(x_3,x_{12},x_{R})$ is
\[\{1, x_3^2x_{12}^2, x_3^2, x_{12}^2, x_3x_{12}\}.\]
Therefore, the Weil descent of  $S_{3,2}(x_3,x_{12},x_{R})$ yields a 
$\n+n$ variable polynomial set $\{f_i\}$ over $\F_2$
such that $\max_i(\deg(f_{i})) = 2$. On the other hand,
the monomial set of $x_3^3\cdot S_{3,2}(x_3,x_{21},x_{R})$ is
\[\{x_3^3,  x_3^5x_{12}^2, x_3^5, x_3^3x_{12}^2, x_3^4x_{12}\}.\]
Therefore, the Weil descent of  $x_3^3\cdot S_{3,2}(x_3,x_{12},x_{R})$ yields a 
polynomial set $\{F_i\}$ over $\F_2$
such that $\max_i(\deg(F_{i})) = 3$. It follows from the definition
that $D_{\mathsf{FirstFall}}(S_{3,2})\le 4$ because the maximum
degree of polynomials obtained from the Weil descent of $x_3^3$ is $2$. 
We conclude that $D_{\mathsf{FirstFall}} \le 4$.

\paragraph{\bf $\boldsymbol {D_{\mathsf{FirstFall}}}$ of the system when $\boldsymbol {m=4}$} 
In this case, one needs to solve the system of $2n$ 
equations over $\F_2$ with $4\n + n$ boolean variables.
The system of equations is derived by applying Weil descent
to (\ref{s: System-4}) that consists of two Semaev polynomials $S_{3,1}$ and $S_{4,2}$. 
From our above discussion, $D_{\mathsf{FirstFall}}(S_{3,1})\le 4$.
Now, analyzing the monomial set of $S_{4,2}(x_3,x_4,x_{123},x_R)$,
we can see that the Weil descent of  $S_{4,2}(x_3,x_4,x_{123},x_R)$ yields a 
$2\n+n$ variable polynomial set $\{f_i\}$ over $\F_2$
such that $\max_i(\deg(f_{i})) = 6$ (this follows from the Weil descent of
the monomial $(x_3x_4x_{123})^3$). On the other hand,
analyzing the monomial set of $x_3\cdot S_{4,2}(x_3,x_4,x_{123},x_R)$,
we see that  the Weil descent of  $x_3\cdot S_{4,2}(x_3,x_4,x_{123},x_R)$ yields a 
polynomial set $\{F_i\}$ over $\F_2$
such that $\max_i(\deg(F_{i})) = 6$.  It follows from the definition
that $D_{\mathsf{FirstFall}}(S_{4,2})\le 7$ because the maximum
degree of polynomials obtained from the Weil descent of $x_3$ is $1$. 
We conclude that $D_{\mathsf{FirstFall}} \le 7$.

\paragraph{\bf $\boldsymbol {D_{\mathsf{FirstFall}}}$ of the system when $\boldsymbol {m=5}$} 
In this case, one needs to solve the system of $2n$ 
equations over $\F_2$ with $5\n + n$ boolean variables.
The system of equations is derived by applying Weil descent
to (\ref{s: System-5}) that consists of two Semaev polynomials $S_{4,1}$ and $S_{4,2}$. 
From our above discussion, $D_{\mathsf{FirstFall}}(S_{4,2})\le 7$.
Now, analyzing the monomial set of $S_{4,1}(x_1,x_2,x_3,x_{123})$,
we can see that the Weil descent of  $S_{4,1}(x_1,x_2,x_3,x_{123})$ yields a 
$3\n+n$ variable polynomial set $\{f_i\}$ over $\F_2$
such that $\max_i(\deg(f_{i})) = 8$ (this follows from the Weil descent of
the monomial $(x_1x_2x_3x_{123})^3$). On the other hand,
analyzing the monomial set of $x_3\cdot S_{4,1}(x_1,x_2,x_3,x_{123})$,
we see that  the Weil descent of  $x_3\cdot S_{4,1}(x_1,x_2,x_3,x_{123})$ yields a 
polynomial set $\{F_i\}$ over $\F_2$
such that $\max_i(\deg(F_{i})) = 8$.  It follows from the definition
that $D_{\mathsf{FirstFall}}(S_{4,1})\le 9$ because the maximum
degree of polynomials obtained from the Weil descent of $x_3$ is $1$. 
We conclude that $D_{\mathsf{FirstFall}} \le 9$.

\section{Experimental results}
\label{s: Experiments}
We implemented the proposed methods in Section~\ref{s: Method} on a 
desktop computer (Intel(R) Xeon(R) CPU E31240 \@ 3.30GHz) using 
Groebner basis algorithms in Magma \cite{Magma}. 
For each parameter set $(n,m,\n)$, we solved 5 random instances of
PDP over a randomly chosen elliptic curve $E/\F_{2^n}$.
In Table~\ref{s: exp results}, we report on our experimental
results for solving PDP($n,m,\n=\lfloor n/m\rfloor$) with $m=3,4,5$.    
In particular, for each of these 5 computations, we report on the maximum CPU time (seconds) and memory (MB)
required for solving PDP. We also report on the maximum of the maximum step degrees $D$
(for which ) in the Groebner basis
computations. Recall that in Section~\ref{s: Method}, we estimated $D_{\mathsf{FirstFall}} \le 4$ when $m=3$;
$D_{\mathsf{FirstFall}} \le 7$ when $m=4$; and $D_{\mathsf{FirstFall}} \le 9$ when $m=5$.
In our experiments, we observe that $D_{\mathsf{reg}} \approx D_{\mathsf{FirstFall}}$.

\begin{table}[h]
\caption{Experimental results on solving PDP($n,m,\n=\lfloor n/m\rfloor$). Time in seconds; Memory in MB; $D$ is
the maximum step degree. }
\label{s: exp results}
\begin{tabular}{|c|c|c|c|c|c|c|c|c|c|}
\cline{1-10}
& \multicolumn{3}{c|}{$m=3$} & \multicolumn{3}{c|}{$m=4$} & \multicolumn{3}{c|}{$m=5$}\\
\cline{1-10}
n & Time &Memory & $D$& Time & Memory & $D$& Time & Memory & $D$\\
\cline{1-10}
11 &&&&&&& 0.520 & 25.8 & 7 \\
12 &&&&&&& 0.670 & 33.0 & 7 \\
13 &&&&&&& 0.890 & 42.8 & 7 \\
14 &&&&&&& 4.260 & 126.7 & 8 \\
15 &&&&&&& 350.100 & 1839.5 & 8 \\
16 &&&& 414.320 & 5100.7 & 7 & 408.270 & 2633.9 & 8 \\
17 & 1.690 & 38.8 & 4 & 1395.170 & 5632.8 & 7 & 506.340 & 4050.3 & 8 \\
18 & 26.680 & 264.5 & 4 &  497.770 & 5632.8 & 7 & 920.790 & 6186.9 & 8 \\
19 & 15.270 & 321.8 & 4 & 509.330 & 5634.1 & 7 & 1265.090 & 8282.9 & 8\\
20 & 49.350 & 397.6 & 4 &  &  & &&&\\
21 & 163.100 & 1228.3 & 4 &  &  &  &&&\\
22 & 126.290 & 1413.2 & 4 &  &  & &&&\\
23 & 248.820 & 1668.7 & 4 &  &  & &&&\\
24 & 1266.610 & 5142.2 & 4 &  & & &&&\\
25 & 1623.180 & 6363.8 & 4 &  &  & &&&\\
26 & 1645.78 & 6596.9 & 4 &  &  & &&&\\
\cline{1-10}
\end{tabular}
\end{table}

Let $m=5$ and $\n = \lfloor n/m\rfloor$. 
Based on our experimental data, it is tempting to assume that 
the underlying system of polynomial equations has 
$D_{\mathsf{reg}}\approx 9$. Moreover,
the system has $N=5\n+n \approx 2n$ boolean variables.
Therefore, when $m=5$, we may estimate the cost of solving ECDLP($2,n$) (see (\ref{s: cost ECDLP}))
as
\begin{align*}
&2^{\n}\frac{2^n m!}{2^{m \n}}{N+D_{\mathsf{reg}}-1 \choose D_{\mathsf{reg}}}^w + 2^{w^\prime \n}\\
&\approx 2^{n/5} m! (2n)^{9w} + 2^{w^\prime n/5}\\
&\approx 2^{34} 2^{n/5} n^{27} + 2^{2n/5},
\end{align*}
where we assume $w=3$ and $w^\prime=2$.
For example, when $n\approx 1200$, the cost of solving ECDLP($2,n$) is estimated
to be $2^{550}$ which is significantly smaller than the cost $2^{600}$ of square-root time algorithms.

\section{Extensions and Optimization}
\label{s: Extend}
In Section~\ref{s: Method}, we introduced a single auxiliary variable to lower the degree
of Semaev polynomials. The degree of polynomials can further be lowered by introducing
more auxiliary variables. As an example, we consider the case $m=5$.
Let $R=(x_R,y_R)\in E$, as before. Notice that there exist $P_i\in \FB$ such that
\[P_1+P_2+P_3+P_4+P_5 - R = \infty\]
if and only if there exist $P_i\in \FB$ and $P_{12},P_{34},P_{50}\in E$ such that
\begin{align}
\label{s: S3 only}
\begin{cases}
P_1+P_2-P_{12} = \infty\\
P_3+P_4-P_{34} = \infty\\
P_5-P_{50}-R = \infty\\
P_{12}+P_{34}+P_{50} = \infty
\end{cases}
\end{align}
Therefore, a $5$-decomposition of $R=P_1+P_2+P_3+P_4+P_5$ may be found as follows:
\begin{enumerate}
\item Define the following system of equations derived from Semaev polynomials
\begin{align}
\label{s: System-5-new}
\begin{cases}
S_{3,1}(x_1,x_2,x_{12}) = 0\\
S_{3,1}(x_3,x_4,x_{34})= 0\\
S_{3,2}(x_5,x_{50},x_R) = 0\\
S_{3,1}(x_{12},x_{34},x_{50})= 0
\end{cases}
\end{align}
Note that this system is defined over $\F_{2^n}$ and has $8$ variables $x_1,x_2,x_3,x_4,x_5,x_{12},x_{34},x_{50}$.
\item Introduce boolean variables $x_{i,j}$ such that 
\[x_i = \sum_{j=0}^{\n -1}x_{i,j}\sigma^j,\] 
for $i=1,2,3,4,5$, and
\[x_{i,j} = \sum_{k=0}^{n}x_{i,j}\sigma^j,\]
for $i=12,34,50$.
Apply the Weil descent technique to (\ref{s: System-5-new}) and define an equivalent system of $4n$ 
equations over $\F_2$ with $5\n + 3n$ boolean variables 
\[\{x_{i,j}:\ i=1,2,3,4,5\ j=0,\ldots\n -1\}\cup\{x_{i,j}:\  i=12,34,50,\ j=0,\ldots n-1\}.\]
Solve this new system of boolean equations and recover $x_1,x_2,x_3,x_4,x_5 \in \F_{2^n}$ from $x_{i,j}\in \F_2$.
\end{enumerate}

Note that the proposed method solves a system of $4n$ 
equations over $\F_2$ with $5\n + 3n$ boolean variables rather
than solving a system of $n$ 
equations over $\F_2$ with $5\n$ boolean variables.
Similar to the analysis in Section~\ref{s: Method}, we can show that
$D_{\mathsf{FirstFall}}\le 4$.

\begin{table}[t]
\caption{Experimental results on solving PDP($n,m,\n=\lfloor n/m\rfloor$). Time in seconds; Memory in MB; $D$ is
the maximum step degree; $D_{\mathsf{Heuristic}}$ is set to be $4$ 
in Groebner basis computations.}
\label{s: exp results 2}
\begin{tabular}{|c|c|c|c|c|c|}
\cline{1-6}
\multicolumn{4}{|c}{} &\multicolumn{2}{|c|}{$D_{\mathsf{Heuristic}}=4$}\\
\cline{1-6}
& \multicolumn{3}{c|}{$m=5$} & \multicolumn{2}{c|}{$m=5$}\\
\cline{1-6}
n & Time &Memory & $D$& Time & Memory\\
\cline{1-6}
11 & 2.380 & 58 & 4 & & \\
12  & 4.150 & 116.7 & 4 & & \\
13 & 6.390 & 124.1 & 4 & & \\
14 & 9.510 & 245.2 & 4 & & \\
15 & 393.170 & 6421.9 & 4 & 7.130 & 256.3\\
16  & 242.500 & 5911.7 & 4 & 6.900 & 320.4\\
17 & 365.460 & 7063.8 & 4 & 6.660 & 320.4\\
18 & 836.080 & 8619.4 & 4 & 11.700 & 394.6\\
19 & 531.420 & 8864.2 & 4 & 45.570 & 2505.3\\
\cline{1-6}
\end{tabular}
\end{table}

In Table~\ref{s: exp results 2}, we report on our experimental
results for solving PDP($n,m,\n=\lfloor n/m\rfloor$) with $m=5$ deploying
only the third Semaev polynomials; see (\ref{s: System-5-new}). 
The time and memory results in the second and third column of 
Table~\ref{s: exp results 2} are obtained using the Groebner basis
implementation of Magma with the $\mathsf{grevlex}$ ordering of monomials.
We observe that the the maximum step degree is $D_{\mathsf{reg}} = 4$ for $11\le n \le 19$. 
The time and memory results in the last two columns of Table~\ref{s: exp results 2} are obtained using 
the Groebner basis implementation of Magma with the $\mathsf{grevlex}$ 
ordering of monomials in a boolean ring. We also introduced two modifications
in the computations: We set the 
$\mathsf{ReductionHeuristic}$ parameter in Magma to $4$; and
we first computed Groebner bases of partial systems described by single equations in (\ref{s: System-5-new}), and merged them later. These two techniques yield non-trivial optimization both in time and memory. For a comparison, when $n=15$ and $m=3$, (Time, Memory) values
decrease from $(393.170, 6421.9)$ to $(7.130, 256.3)$ when this modification is deployed in the computation; see Table~\ref{s: exp results 2}. For the same parameters ($n=15$ and $m=3$),
(Time, Memory) values are reported as ($174.47, 2635.4$) in \cite{Semaev-15}.

Based on our experimental data, we may assume that the underlying system of polynomial equations has 
$D_{\mathsf{reg}}\approx 4$ for all $n$. Moreover,
the system has $N=5\n+3n \approx 4n$ boolean variables.
Therefore, when $m=5$, we may estimate the cost of solving ECDLP($2,n$) (see (\ref{s: cost ECDLP}))
as
\begin{align*}
&2^{\n}\frac{2^n m!}{2^{m \n}}{N+D_{\mathsf{reg}}-1 \choose D_{\mathsf{reg}}}^w + 2^{w^\prime \n}\\
&\approx 2^{n/5} m! (4n)^{4w} + 2^{w^\prime n/5}\\
&\approx 2^{31} 2^{n/5} n^{12} + 2^{2n/5},
\end{align*}
where we assume $w=3$ and $w^\prime=2$.
This running time outperforms square-root methods when $n>457$.
For example, when $n\approx 550$, the cost of solving ECDLP($2,n$) is estimated
to be $2^{250}$ which is significantly smaller than the cost $2^{275}$ of square-root time algorithms.

\section*{Acknowledgment}
I would like to thank Michiel Kosters and Igor Semaev for their comments 
on the first version of this paper.

\bibliographystyle{amsplain}



\end{document}